\documentstyle[11pt,graphicx]{article}

\newcommand{\be}{\begin{equation}}
\newcommand{\ee}{\end{equation}}
\newcommand{\ba}{\begin{eqnarray}}
\newcommand{\ea}{\end{eqnarray}}

\begin{document}

\title{Statistical Model of Superconductivity in a 2D Binary Boson-Fermion
Mixture}
\author{\and M. Casas$^a$, N.J. Davidson$^b$, M. de Llano$^c$, T.A.
Mamedov$^{d}$ \\
A. Puente$^a$, R.M. Quick$^b$, A. Rigo$^a$ and M.A. Sol\'{\i}s$^e$}
\maketitle
\noindent $^{a}$Departament de F\'{\i}sica, Universitat de les Illes
Balears, 07071 Palma de Mallorca, Spain \\[0pt]
$^{b}$Department of Physics, University of Pretoria, 0002 Pretoria, South
Africa\\[0pt]
$^{c}$Instituto de Investigaciones en Materiales, Universidad Nacional
Aut\'{o}noma de M\'{e}xico,\\[0pt]
Apdo. Postal 70-360, 04510 M\'{e}xico, DF, Mexico \\[0pt]
$^{d}$ Department of Engineering, Baskent University, Baglica, 
06530 Ankara, Turkey \\[0pt]
$^{e}$Instituto de F\'{\i}sica, Universidad Nacional Aut\'{o}noma de
M\'{e}xico, 01000 M\'{e}xico, DF, Mexico

\begin{abstract}
A two-dimensional (2D) assembly of noninteracting, temperature-dependent,
composite-boson Cooper pairs (CPs) in chemical and thermal equilibrium with
unpaired fermions is examined in a binary boson-fermion statistical model
as the superconducting singularity temperature is approached from above. The
model is derived from {\it first principles} for the BCS
model interfermion interaction from three extrema of the system Helmholtz
free energy (subject to constant pairable-fermion number) with respect to: a)
the pairable-fermion distribution function; b) the number of excited (bosonic)
CPs, i.e., with nonzero total momenta---usually ignored in BCS theory---and
with
the appropriate (linear, as opposed to quadratic) dispersion relation that
arises from the Fermi sea; and c) the number of CPs with zero total momenta.
Compared with the BCS theory condensate, higher singularity temperatures
for the
Bose-Einstein condensate are obtained in the binary boson-fermion mixture model
which are in rough agreement with empirical critical temperatures for quasi-2D
superconductors. \newline

{\bf Keywords:}  Binary gas; boson-fermion mixture; Bose-Einstein condensation

{\bf PACS numbers:}  03.75Fi;  05.30.-d;  05.30.Fk;  05.30.Jp


{\bf Corresponding author: }\ \ M. de Llano, IIM-UNAM, Apdo. Postal 70-360,
04510 M\'{e}xico, DF, Mexico, Fax: +52-5616-1251,
dellano@servidor.unam.mx
\end{abstract}





\pagebreak

\section{Introduction}

Recent experiments \cite{Fried} indicate that composite bosons in ultra-cold
clouds of most alkali atoms do indeed Bose-Einstein (BE) condense. Since
Cooper pairs (CPs) of fermions (electrons or holes) in a many-fermion system
form composite bosons in the sense of coupling to integer angular momentum,
it is natural to consider the possible BE condensation of such pairs. The
belief that some such condensate is central to superconductivity is more
than 50 years old \cite{Ogg,Ginzburg,Schafroth,Blatt,Anderson,Mu}.
High-$T_{c}$,
as well as some organic, superconductors \cite{Poole} are
quasi-two-dimensional (2D). Quasi-1D superconductors have also been found
\cite{salts}. BE condensation (BEC) is impossible in two or less space
dimensions \cite{Hohenberg} for usual or ``ordinary'' bosons (i.e., with a
quadratic energy-momentum, or dispersion, relation). It is however still
possible to have BEC in all dimensions $d>1$ for non-interacting bosons if
they obey a {\it linear} dispersion relation \cite{Fujita1}---such as CPs
moving in the Fermi sea. This possibility arises because the Hohenberg
theorem \cite{Hohenberg}, which prohibits BEC in 2D, relies on an $f$-sum
rule based on the quadratic dispersion relation appropriate to bosons \cite
{Nozieres} moving in a vacuum. \ Such a linear dispersion relation for the
CPs in a binary boson-fermion mixture model was recently found \cite{FM}\ to
be consistent, without any adjustable parameters, with the anomalous linear
(quadratic) temperature-dependence above $T_{c}$ in the resistivity of
optimally-doped (overdoped) cuprates whether hole- or electron-doped. \ For
the observed quadratic $T$-dependence in overdoped samples linear-dispersion
CP charge carriers are {\it essential}.

Although extensive studies in the BCS-Bose ``crossover'' problem in
superconductivity have spanned \cite{crossover} a period of over thirty
years, we note that BEC is distinct from the standard (i.e., zero
center-of-mass
momentum CPs) BCS theory condensation where only that {\it one} bosonic state
exists.

In this paper it is shown that in addition BEC is still possible in 2D even if
the number of composite bosons (pairs of fermions) in a binary boson-fermion
mixture is not fixed---as chemical/thermal equilibrium renders it coupling- and
temperature-dependent---as long as the total number of fermions is fixed.
This gives rise to an interesting statistical-mechanics problem irrespective
of the particular mechanism for pair formation, and may have a vital
application for superconductivity as well as for (neutral-atom)
superfluidity such as in liquid $^{3}$He \cite{VW}, dilute mixtures of
$^{3}$He
in $^{4}$He \cite{Edwards}, or in trapped Fermi gases
\cite{tfg}. \ The statistical model dealt with here may be seamlessly linked
to BCS theory, via the fermionic energy gap, when boson/unpaired-fermion
interactions are included as, e.g., in
Refs. \cite{TDLee} and \cite{Tolma}.  However, in these two papers the
quadratic CP dispersion relation has been assumed. \ The quadratic form has
recently been shown\ \cite{Adh}\ to apply {\it only }in the zero-density or
vacuum limit when the Fermi sea disappears.

In Section 2 we recall a 2D gas of fermions at  $T=0$ interacting
via a constant pairing interaction in an annulus about the Fermi
surface---viz.,
the {\it BCS model interaction}. The binding energy of a {\it single} pair
near the Fermi surface (CP problem) decreases practically linearly with the
center-of-mass momentum (CMM) of the pair for all values of the momentum
below breakup, the breakup momentum typically being only about four orders
of magnitude smaller than the Fermi momentum. In Section 3 we discuss why the
interacting many-fermion system can be
treated as a set of independent CPs (i.e., composite bosons with fermion
number two) mixed in with pairable fermions which are not bound into pairs,
i.e., unpaired fermions. In Section 4 the more realistic scenario is
considered
of the BEC of these pairs, incorporating pair breakup beyond a certain CMM.
Although the number of pairs is not fixed but rather strongly coupling- and
temperature-dependent, BEC is still possible in 2D. A simple binary boson-
fermion statistical model is introduced by constructing the Helmholtz free
energy for an ideal mixture of pairable but unpaired fermions plus paired
fermions (both zero and nonzero CMM pairs), all in chemical and thermal
equilibrium. The latter results through extrema of the free energy in: \ a)
the pairable fermion occupation probabilities; b) the excited boson numbers
(nonzero CMM CPs); and c) the ground boson number (zero CMM pairs).  In
Section 5 the coupling- and temperature-dependence of the boson number is
derived. In Section 6 the critical BEC singularity temperature is obtained
first by ignoring the unpaired fermions in a pure boson-gas model and then
exactly for the boson-fermion binary mixture model from a $T$-dependent
dispersion relation derived and calculated numerically, and results compared
with empirical data.  Finally, Section 7 gives conclusions.

\section{Cooper-pair dispersion relation}

Consider a 2D system of N fermions of mass $m$ confined in a square ``pen''
of area $L^{2}$ and interacting pairwise via the BCS model interaction
\begin{equation}
V_{{\bf k},{\bf k}^{\prime }}=\left\{
\begin{array}{cl}
-V & \mbox{if}\;\;\mu (T)-\hbar \omega _{D}<\epsilon _{k_{1}}(\equiv
\hbar ^{2}k_{1}^{2}/2m),\ \epsilon _{k_{2}}<\mu (T)+\hbar \omega _{D} \\
0 & \mbox{otherwise},
\end{array}
\right.   \label{inter}
\end{equation}
where ${\bf k}\equiv {\frac{1}{2}}({\bf k}_{1}-{\bf k}_{2})$ is the relative
wavevector of the two particles (see Fig. 3 below); $V_{{\bf k},{\bf k}^{\prime }}$ the 2D
double Fourier integral of the underlying non-local interaction $V({\bf r},%
{\bf r}^{\prime })$ in the relative coordinate ${\bf r}={\bf r}_{1}-{\bf r}%
_{2};\ \ \mu (T)$ the ideal Fermi gas chemical potential which at $T=0$
becomes the Fermi energy $E_{F}\equiv \hbar ^{2}k_{F}^{2}/2m$ with $k_{F}$
the Fermi wavenumber; $\hbar \omega _{D}\equiv \hbar ^{2}k_{D}^{2}/2m$ the
width of the annulus about the Fermi circle in which the pairing interaction
is nonzero, with $\omega _{D}$ being the Debye frequency. This model
interaction mimics the net effect of an attractive electron-phonon
interaction overwhelming the repulsive interfermion Coulomb repulsions,
whenever $V>0$.

If $\hbar {\bf K}=\hbar ({\bf k}_{1}+{\bf k}_{2})$ is the center-of-mass
momentum (CMM) of a pair, let $E_{K}$ be its {\it total} energy (besides the
CP rest-mass energy). The eigenvalue (CP \cite{Coo}) equation for a pair of
fermions at $T=0$
immersed in a background of $N-2$ inert, spectator fermions within a (sharp)
Fermi circular perimeter of radius $k_{F}$ is then
\begin{equation}
1=V{\sum_{{\bf k}}}^{^{\prime }}\frac{\theta (k_{1}-k_{F})\,\,\theta
(k_{2}-k_{F})}{2\epsilon _{k}-(E_{K}-{\hbar ^{2}}K{^{2}/}4m)},
\label{eq:cooper}
\end{equation}
where again $\epsilon _{k}\equiv \hbar ^{2}k^{2}/2m$, $\theta (x)$ is the
Heaviside unit step function, and the prime on the summation sign denotes
the conditions

\begin{equation}
k_{1}\equiv |{\bf k}+{\frac{1}{2}}{\bf K}|<(k_{F}^{2}+k_{D}^{2})^{1/2}\;\;\;%
\mbox{and}\;\;\;k_{2}\equiv |{\bf k}-{\frac{1}{2}}{\bf K}%
|<(k_{F}^{2}+k_{D}^{2})^{1/2}  \label{limits}
\end{equation}
ensuring that the pair of fermions {\it above} the Fermi ``surface'' cease
interacting beyond the annulus of energy thickness $\hbar \omega _{D}$ in
accordance with (\ref{inter}), thereby restricting the summation over ${\bf k%
}$ for a given fixed ${\bf K}$. Without these restrictions (\ref{eq:cooper})
would just be the Schr\"{o}dinger equation in momentum space for the pair.
Setting $E_{K}\equiv 2E_{F}-\Delta _{K}$, the pair is {\it bound} if $\Delta
_{K}>0$, and (\ref{eq:cooper}) becomes an eigenvalue equation for the
(positive) pair binding energy $\Delta _{K}$. Our $\Delta _{K}$ and $\Delta
_{0}$
should {\it not} be confused with the BCS energy gap $\Delta(T)$.

Let $%
\lambda \equiv g(E_{F})V\geq 0$ be a dimensionless coupling constant with $%
g(E_{F})$ the electronic density-of-states (for each spin) at the Fermi
surface in the normal (i.e., interactionless) state, which in 2D is constant

\begin{equation}
g(\epsilon )=L^{2}m/2\pi \hbar ^{2} \equiv g.  \label{gdee}
\end{equation}
The Cooper equation (\ref{eq:cooper}) for the unknown quantity $\Delta_{K}$ is
analyzed in Ref. \cite{PhysC}. For zero CMM, $K=0$, it becomes
a single elementary integral, with the familiar \cite{Coo} solution
\begin{equation}
\Delta_{0}=\frac{2\hbar \omega_{D} }{e^{2/\lambda }-1}  \label{delta0}
\end{equation}
valid for {\it all} coupling $\lambda $. For small $K$, it is not too
difficult to extract \cite{PhysC} the asymptotic result
\begin{equation}
{\Delta _{K}}\mathrel{\mathop{\longrightarrow}\limits_{K \rightarrow 0}}%
\Delta _{0}-{\frac{2}{\pi }}\big [1+{\frac{\Delta _{0}}{2\hbar \omega _{D}}}%
(1+\sqrt{1+\nu })\big ]\hbar v_{F}K+O(K^{2})\mathrel{\mathop{%
\longrightarrow}\limits_{\lambda \rightarrow 0}}\Delta _{0}-{\frac{2}{\pi }}%
\hbar v_{F}K+O(K^{2})  \label{eq:linear}
\end{equation}
where $\nu \equiv \Theta _{D}/T_{F}$, and $v_{F}$ is the Fermi velocity
defined through $E_{F}\equiv \hbar
^{2}k_{F}^{2}/2m={\frac{1}{2}}mv_{F}^{2}$. For weak coupling, $\lambda
\rightarrow 0$, this {\it linear dispersion relation} gives the 2D analog of
the 3D result stated as far back as 1964 in Ref. \cite{Schrieffer}, p. 33
(see also, Ref. \cite{FW}, p. 336) but with the 2D coefficient $2/ \pi$
of the last expression of (\ref {eq:linear}) replaced by $1/2$.

\section{Justification of boson formalism}

These CP boson-like structures could be called ``quasi-bosons'' since their
creation and annihilation operators are known {\it not} to obey the usual
boson commutation relations \cite{Schrieffer}, p. 38. However, they do obey
the Bose-Einstein distribution since the energy $E_{K}$ of the CP is given
only by the total CMM, $K$, but is {\it independent} of the relative momentum
$k$.  Thus, the possible energy states for the pair are $E_{K}$ as defined in
(\ref {eq:cooper}). The number of pairs $N_{{\bf K}}$ that can occupy such a
state can take on indefinite values since there exist also indefinitely many
relative momenta, namely
\begin{equation}
N_{{\rm {\bf K}}}\equiv \sum_{{\rm {\bf k}}}{\cal N}_{{\rm {\bf k,K}}%
}=0,1,2,...\,.
\end{equation}
Here, ${\cal N}_{{\rm {\bf k,K}}}=0,1$ is the number of pairs characterized
by both {\bf k} and {\bf K}, and is the same number as that characterized by
definite ${\rm {\bf k}}_{1}$ and ${\rm {\bf k}}_{2}$, namely ${\cal N}_{{\rm
{\bf k,K}}}=n_{{\rm {\bf k}}_{1}}n_{{\rm {\bf k}}_{2}}=0,1$ where $n_{{\rm
{\bf k}}_{i}}=0,1$ is the occupation number for a {\em single} fermion,
these remarks all referring to {\em singlet} pairing. Much of all this has been
known \cite{Matsu} at least since 1958, albeit in somewhat different
language.

This view of an {\it actual} Cooper {\it pair} should not be confused with,
say, an Anderson \cite{An} phonon-like collective {\it excitation} (or {\it %
modes}) with weak-coupling dispersion relation---in 2D \cite{BR} given by $%
(1/\sqrt{2})\hbar v_{F}K$ in the long-wavelength limit, and which evolves
into the {\it plasmon} when Coulomb repulsions between fermions are switched
on. CPs here, like deuterons, carry fermion number two and as such are {\it %
definite} in number (although in the CP case this number is coupling- and
temperature-dependent) and can thus undergo BEC. This is distinct from
collective excitations which are indefinite in number. Park \cite{Park},
e.g., distinguishes between ``permanent'' and ``ephemeral'' bosons, the
latter sometimes being referred to as ``quasiparticles'' to distinguish from
the former ``particles''.

For $N_{B}$ ordinary bosons of mass $m_{B}$ and energy $\varepsilon
_{K}=\hbar ^{2}K^{2}/2m_{B}$ in any positive dimension, $d>0$, a temperature
singularity $T_{c}$ \cite{Gunton} appears in the number equation $%
N_{B}=\sum_{{\bf K}}[e^{(\varepsilon _{K}-\mu _{B})/k_{B}T}-1]^{-1}$ at
vanishing bosonic chemical potential $\mu _{B}%
\mathrel{\mathop{\textstyle
<}\limits_{\sim }}0$ when the number of ${\bf K}=0$ bosons just ceases to be
negligible upon cooling. It is given by
\begin{equation}
T_{c}={\frac{2\pi \hbar ^{2}}{m_{B}k_{B}}}\bigg[{\frac{n_{B}}{g_{d/2}(1)}}%
\bigg]^{2/d}  \label{eq:becmnz}
\end{equation}
with $n_{B}$ the boson particle density $N_{B}/L^{d}$, and $g_{d/2}(z)$ the
usual Bose integrals
\begin{equation}
g_{\sigma }(z)\equiv \frac{1}{\Gamma (\sigma )}\int_{0}^{\infty }dx\frac{%
x^{\sigma -1}}{z^{-1}e^{x}-1}=\sum_{l=1}^{\infty }\frac{z^{l}}{l^{\sigma }}%
\mathrel{\mathop{\longrightarrow}\limits_{ z \rightarrow 1}}\zeta (\sigma ),
\label{gsigma}
\end{equation}
where $\Gamma (\sigma )$ is the gamma function and $\zeta (\sigma )$ the
Riemann zeta function of order $\sigma $. The last identification in (\ref
{gsigma}) holds when $\sigma >1$ for which $\zeta (\sigma )<\infty $,
while the series $g_{\sigma }(1)$ diverges for $\sigma \leq 1$, thus
giving $T_{c}=0$ for $d\leq 2$. For $d=3$ one has $\zeta (3/2)\simeq 2.612$
so that (\ref{eq:becmnz}) becomes the familiar formula $T_{c}\simeq
3.31\hbar ^{2}n_{B}^{2/3}/m_{B}k_{B}$ of ``ordinary'' BEC. On the other
hand, for bosons with (positive) excitation energy $\varepsilon _{K}\equiv
\Delta _{0}-\Delta _{K}$ given approximately by the linear term in (\ref
{eq:linear}) for all $K$, the singularity that lead to (\ref{eq:becmnz}) now
yields \cite{VCAN}, for weak coupling,
\begin{equation}
T_{c}=\frac{a(d)\hbar v_{F}}{k_{B}}\bigg[{\frac{\pi ^{\frac{d+1}{2}}n_{B}}{%
\Gamma ({\frac{d+1}{2}})g_{d}(1)}}\bigg]^{1/d}  \label{eq:becmz}
\end{equation}
where \cite{Fujita1} $a(d)=1,\ \ 2/\pi $ and $1/2$ for $d=1,\ \ 2$ and $3$,
respectively. Note that now ${T_{c}>0}$ {\it for all} ${d>1}$, which is {\it %
precisely} the dimensionality range of all known superconductors including
the quasi-1D organo-metallic (Bechgaard) salts \cite{salts}. This is {\it not%
} inconsistent with the Hohenberg theorem \cite{Hohenberg} that there is no
broken symmetry, i.e., long-range order, in a Bose fluid for $d$ = 1 or 2,
since this is based on an $f$-sum rule for bosons with a quadratic
dispersion relation. Indeed, both (\ref{eq:becmnz}) and (\ref{eq:becmz}) are
special cases of of the more general expression \cite{cas}\ for any space
dimensionality $d>0$ and any boson dispersion relation $\varepsilon
_{K}=C_{s}\,K^{s}$ with $s$\ $>0$ and $C_{s}$ a constant, given by
\begin{equation}
T_{c}=\frac{C_{s}}{k_{B}}\left[ \frac{s\,\Gamma (d/2)\,(2\pi )^{d}n_{B}}{%
2\pi ^{d/2}\,\Gamma (d/s)g_{d/s}(1)}\right] ^{s/d}.  \label{gentc}
\end{equation}

In what follows the number of bosons will be temperature-dependent and it is
{\it in conserving the fermion number} that the singularity arises. As is the
case for the pure boson gas, a linear rather than a quadratic dispersion
relation will be needed to obtain BEC in 2D. This emerges in a statistical
model
of an ideal binary {\it mixture} of bosons (the CPs) and unpaired (both
pairable and unpairable) fermions in chemical equilibrium \cite{Schafroth},
for which thermal pair-breaking into unpaired pairable fermions is
explicitly allowed.

\section{First-principles statistical model}

Under interaction (\ref{inter}) at any $T$ the total number of fermions in
2D is $N=L^{2}k_{F}^{2}/2\pi =N_{1}+N_{2}$ and is just the number of
non-interacting (i.e., unpairable) fermions $N_{1}$ plus the number of
pairable ones $N_{2}$. The unpairable fermions obey the usual Fermi-Dirac
distribution with fermionic chemical potential $\mu $. On the other hand,
the $N_{2}$ pairable fermions are simply those in the interaction shell of
energy width $\hbar \omega _{D}$ so that
\begin{equation}
N_{2}=2\int_{\mu -\hbar \omega _{D}}^{\mu +\hbar \omega _{D}}\;d\epsilon
\frac{%
 g(\epsilon) }{e^{\beta (\epsilon -\mu )}+1}=2g\hbar \omega _{D},
 \label{n214}
\end{equation}
since the density of electronic states (\ref{gdee}) is constant and the
remaining integral exact. At any interfermionic coupling and temperature
these fermions form an ideal mixture of pairable but unpaired fermions plus
CPs that are created near the single-fermion energy $\mu (T)$, with binding
energy $\Delta _{K}(T)$ $\geq 0$ and total energy
\begin{equation}
E_{K}(T)\equiv 2\mu (T)-\Delta _{K}(T).  \label{bosonenergy}
\end{equation}
This generalizes the $T=0$ equation $E_{K}\equiv 2
E_{F}-\Delta _{K}$ introduced below (\ref{limits}).

The Helmholtz free energy $F=E-TS$, where $E$ is the internal energy and $S$
the entropy, for this binary {\it ``composite
boson/pairable-but-unpaired-fermion system''} at temperatures $T\leq T_{c}$
is then \cite{Path}
\begin{eqnarray}
F_{2} &=&2\int_{\mu -\hbar \omega _{D}}^{\mu +\hbar \omega _{D}}d\epsilon
\;g(\epsilon )\bigg\{n_{2}(\epsilon )\epsilon +k_{B}T\left[
n_{2}(\epsilon )\ln n_{2}(\epsilon )+\{1-n_{2}(\epsilon )\}\ln
\{1-n_{2}(\epsilon )\}\right] \bigg\}  \nonumber \\
&&+[2\mu (T)-\Delta _{0}(T)]N_{B,0}(T)  \nonumber \\
&&+\sum_{K>0}^{K_{0}}\bigg\{[2\mu (T)-\Delta _{K}(T)]N_{B,K}(T)  \nonumber \\
&&+k_{B}T\left[ N_{B,K}(T)\ln N_{B,K}(T)-\{1+N_{B,K}(T)\}\ln \{1+N_{B,K}(T)\}%
\right] \bigg\}.  \label{eq:f2}
\end{eqnarray}
The integral term is the contribution from the unpaired fermions and runs
over all levels in the energy shell where the BCS model interaction is
nonzero, $n_{2}(\epsilon )$ being the average number of unpaired but
pairable fermions with energy $\epsilon $; the prefactor two comes from
the spin. The second term gives the free energy of the bosons with CMM $K=0$
since their entropy is negligible in the thermodynamic limit; here $%
N_{B,0}(T)$ is the number of (bosonic) CPs with zero CMM at temperature $T$.
The summation term represents the free energy of the bosons with nonzero
CMM, while $N_{B,K}(T)$ is that with arbitrary nonzero CMM $K,$ and the
cutoff $K_{0}$ is defined \cite{PhysC} by $\Delta _{K_{0}}\equiv 0$. The free
energy $F_{2}$ is to be minimized subject to the constraint that the total
number of pairable fermions $N_2$ is conserved.

If $N_{20}(T)$ is the number of pairable but unpaired fermions, the relevant
{\it number equation} for the pairable (i.e., active) fermions is then
\begin{equation}
N_{2}=N_{20}(T)+2[N_{B,0}(T)+N_{B,0<K<K_{0}}(T)]\equiv N_{20}(T)+2N_{B}(T),
\label{eq:number}
\end{equation}
where $N_{B,0<K<K_{0}}(T)$ denotes the {\it total} number of ``excited''
bosonic pairs (namely with CMM such that $0<K<K_{0}$), i.e., $%
N_{B,0<K<K_{0}}(T)\equiv \sum_{0<K<K_{0}}N_{B,K}(T)$. Minimizing the free
energy, subject to the constraint that (\ref{eq:number}) be a constant, is
equivalent to minimizing the grand potential
\begin{equation}
\Omega _{2}=F_{2}-\mu _{2}N_{2}.  \label{eq:omega}
\end{equation}

\hspace{0.5cm} {\bf a)} Minimizing $\Omega _{2}$ with respect to the {\bf %
fermion occupation probabilities} $n_2(\epsilon)$ yields the Fermi-Dirac
distribution with fermion chemical potential $\mu _{2}$, {\it not} $\mu$,
namely

\begin{equation}
n_{2}(\epsilon )=\frac{1}{e^{\beta (\epsilon -\mu _{2})}+1}\ ;\ \ \ \ \
\beta \equiv (k_{B}T)^{-1}.
\end{equation}
Thus the total number of pairable (but unpaired) fermions then becomes

\begin{equation}
N_{20}(T) \equiv 2\int_{\mu -\hbar \omega _{D}}^{\mu +\hbar \omega
_{D}}d\epsilon
\;g(\epsilon ) n_{2}(\epsilon )=2\int_{\mu -\hbar \omega _{D}}^{\mu +\hbar
\omega
_{D}}d\epsilon \,\,{\frac{g(\epsilon )}{e^{\beta (\epsilon -\mu
_{2})}+1}}\ ,  \label{n20t}
\end{equation}
and should be compared with (\ref{n214}) for $N_{2}$ which contains only $%
\mu $. Since in 2D $g(\epsilon )$ is a constant (\ref{gdee}), (\ref{n20t}%
) becomes the exact expression

\begin{equation}
N_{20}(T)={\frac{2g}{\beta }}\ln \left[ {\frac{1+e^{-\beta (\mu -\mu
_{2}-\hbar \omega _{D})}}{1+e^{-\beta (\mu -\mu _{2}+\hbar \omega _{D})}}}%
\right] .  \label{eq:unpaired}
\end{equation}

\hspace{0.5cm} {\bf b)} Minimizing $\Omega _{2}$ with respect to the {\bf %
excited boson numbers} $N_{B,K}(T)$, $K>0$, yields the Bose-Einstein
distribution summed over all $0<K<K_{0}$, namely

\begin{equation}
N_{B,0<K<K_{0}}(T)\equiv
\sum_{K>0}^{K_{0}}N_{B,K}(T)=\sum_{K>0}^{K_{0}}[e^{\beta \{E_{K}(T)-2\mu
_{2}\}}-1]^{-1}.  \label{eq:exbosons}
\end{equation}
The factor multiplying $\beta $ in (\ref{eq:exbosons}) may be rewritten as $%
\varepsilon _{K}{(T)}-\mu _{B}{(T)}$, where $\varepsilon _{K}{(T)}%
\equiv \Delta _{0}{(T)}-\Delta _{K}{(T)}\geq 0$ is a (nonnegative) excitation
energy as suggested by (\ref{eq:linear}), while $\mu _{B}{(T)}$ turns out to be

\begin{equation}
\mu _{B}(T)=2\left[ \mu _{2}(T)-\mu (T)\right] +\Delta _{0}(T).  \label{mub}
\end{equation}
This allows rewriting (\ref{eq:exbosons}) in the more meaningful {\it boson}
form
\begin{equation}
N_{B,0<K<K_{0}}(T)=\sum_{K>0}^{K_{0}}[e^{\beta \{\varepsilon
_{K}(T)-\mu _{B}(T)\}}-1]^{-1}
\label{N_BK}
\end{equation}
where $\mu _{B}(T)$ is clearly the bosonic chemical potential associated
with the entire binary mixture.

\hspace{0.5cm} {\bf c)} Finally, minimizing $\Omega _{2}$ with respect to
the {\bf number of zero CMM (or, \lq\lq ground state") bosons $N_{B,0}(T)$}
gives

\begin{equation}
2[\mu _{2}(T)-\mu (T)]+\Delta _{0}(T)=0\;\;\;\;\;\;(0\leq T\leq T_{c}),
\label{eq:chemeqb}
\end{equation}
valid only in the stated temperature range as $N_{B,0}(T)$ is negligible for
all $T>T_{c}$. However, in view of \ (\ref{mub}) this implies that $\mu
_{B}(T)$ $=0$ for all $0\leq T\leq T_{c}$---which is precisely the BEC
condition for a {\it pure }boson gas, even though one now deals with a
binary boson-fermion {\it mixture}. \

\section{Boson number}
To determine $N_{B}(T)$ from (\ref{eq:number}) we need (\ref{eq:unpaired})
which
with (\ref{eq:chemeqb}) reduces to

\begin{equation}
N_{20}(T)={\frac{2g}{\beta }}\ln \left[ {\frac{{1+e^{-\beta \{\Delta
_{0}(T)/2-\hbar \omega _{D} \} }}}{{1+e^{-\beta \{\Delta _{0}(T)/2+\hbar \omega
_{D} \} }}}}\right] \hspace{0.5in}(0\leq T\leq T_{c}).  \label{n20t26}
\end{equation}
At $T=0$ two distinct coupling regimes emerge by inspecting (\ref{n20t26}): a)
for $\Delta _{0}/2<\hbar \omega _{D}$ or, from (\ref{delta0}) for
$\lambda \leq 2/ \ln 2 \simeq 2.89$,
we have that $N_{20}(0)=2g(\mu )(\hbar \omega _{D}-\Delta_{0}/2)$;
while b) for $\Delta _{0}/2>\hbar \omega _{D}$ (or $\lambda \geq 2.89$)
$N_{20}(0)$ is identically zero. Hence, the number of bosons $%
N_{B}(0) $ at $T=0$\ from (\ref{eq:number}) is just $N_{B}(0)={\frac{1}{2}}%
[N_{2}-N_{20}(0)]$. Using (\ref{n214}) for $N_{2}$ the {\it fractional
number of
pairable fermions that are actually paired} at $T=0$, namely $%
2N_{B}(0)/N_{2}=1-N_{20}(0)/N_{2}$, becomes simply

\begin{equation}
2N_{B}(0)/N_{2}=\left\{
\begin{array}{ll}
\Delta _{0}/2\hbar \omega _{D}=(e^{2/\lambda }-1)^{-1}\mathrel{\mathop{%
\longrightarrow}\limits_{\lambda \rightarrow 0}}e^{-2/\lambda } & ({\rm for}%
\ \lambda \leq 2/\ln 2\simeq 2.89) \\
1 & ({\rm for}\ \lambda \geq 2/\ln 2\simeq 2.89).
\end{array}
\right.  \nonumber  \label{NB/N2}
\end{equation}
This fraction is plotted against coupling $\lambda $ in Fig. 1 as $2n_B(0)/n_2$, since $n_B(T) \equiv N_B(T)/L^2$ and $n_2 \equiv N_2/L^2$. Since $%
N_{B}(0)={\frac{1}{2}}g\Delta _{0}$ for $\lambda \leq 2.89$, only
those fermions in an energy shell of width $\Delta _{0}/2$ around the Fermi
surface actually pair at $T=0$, while for $\lambda \geq 2.89$ {\it all}
pairable fermions actually pair up since then $N_{B}(0)=g\hbar \omega
_{D}\equiv {\frac{1}{2}}N_{2}$. This result contrasts sharply with the
``heuristic model'' \cite{cas}, Eq. (\ref{N_BK}), where $2N_{B}(0)/N_{2}$
$\equiv 1$ for {\it all }coupling, and is more in line with BCS theory
which implies, in any $d$,
a coupling-dependent fraction estimated (Ref. \cite{Blatt} p. 128; see also
\cite {Koch}) to be $[g(E_{F})2\Delta /2g(E_{F})\hbar \omega
_{D}]^{2}= (\Delta /\hbar \omega
_{D})^{2} \equiv (\sinh 1/\lambda )^{-2}%
\mathrel{\mathop{\longrightarrow}\limits_{\lambda
\rightarrow 0}}4e^{-2/\lambda }$, where $\Delta \equiv \hbar \omega
_{D}/\sinh (1/\lambda )$ (again, not to be confused with the CP binding energy
$\Delta _{0}$) is the $T=0$ BCS energy gap for the same BCS model interaction
(\ref{inter}) used in this paper; this is graphed as the thin curve
in Fig. 1 and is seen to be much larger than (\ref{NB/N2}) for fixed $%
\lambda $. The breakdown of BCS theory for BCS model interaction couplings
larger than $\lambda \simeq 1.13$ is clear both because:  a) the alluded
fraction cannot exceed unity; and b) physically, if the fermionic energy
gap $\Delta \geq \hbar \omega _{D}$ no pairable fermions are available at all.
This breakdown is indicated by the dashed curve in Fig. 1. (A
strong-coupling many-body model differing from that of BCS theory but based on
the BCS model interaction has been solved by Thouless \cite{thou}).\

Also displayed in Fig. 1 are two finite-temperature
results for $2N_{B}(T)/N_{2}=1-N_{20}(T)/N_{2}$ which are obtainable from
(\ref{n20t26}) for any $T$ provided one knows $\Delta_{0}(T)$ for any $T > 0$.
For $T>0$, the $\theta(k_{1}-k_{F})\equiv \theta (\epsilon _{k_{1}}-E_{F})$
in (\ref{eq:cooper}) becomes $1-n(\xi _{k_{1}})$, where $n(\xi
_{k_{1}})\equiv (e^{\beta \xi _{k_{1}}}+1)^{-1}$ is the Fermi-Dirac
distribution with $\xi _{k_{1}}\equiv
\epsilon _{k_{1}}-\mu (T)$, with the ideal fermion gas chemical potential
$\mu (T)$ in 2D being given exactly by
\begin{equation}
\mu (T)=\beta ^{-1}\ln (e^{\beta E_{F}}-1)\mathrel{\mathop {\longrightarrow
}\limits_{T\rightarrow 0}}E_{F}.  \label{muT}
\end{equation}
Note that $\mu (T)$ decreases monotonically with temperature from its
maximum value of $E_{F}$ but does not turn negative until $%
T=T_{F}/\ln 2\simeq 1.44T_{F}$ so that the BCS model interaction (%
\ref{inter}), which {\it requires }$\mu (T)$ to be nonnegative, will not
break down (i.e., become meaningless) over the entire range of temperatures
relevant in this paper, see Fig. 4 below.\ Similar arguments hold for
$\theta (k_{2}-k_{F})$.
Since $k_{1}=k_{2}$ implies that $\xi _{k_{1}}=\xi _{k_{2}}$, (\ref
{eq:cooper}) then leads to a simple generalization to finite-temperature of
the
$K=0$ CP equation, namely
\begin{equation}
1=\lambda \int_{0}^{\hbar \omega _{D}}d\xi (e^{-\beta \xi }+1)^{-2}[2\xi
+\Delta _{0}(T)]^{-1}.  \label{e13}
\end{equation}
Its numerical solution for $\Delta _{0}(T)$ is illustrated in Fig. 2 for $%
\lambda \equiv gV =1/2$ and $\nu \equiv \hbar \omega _{D}/E_{F}=0.05$. Note
that
if one assumes a $T^*$ such that $\Delta_0(T^*)=0$, the resulting integral in
(\ref {e13}) diverges and the equation can only be satisfied for $\lambda =0$;
thus, there is no temperature $T^*$ at which \lq \lq depairing" will occur for
any fixed $\nu$ and any nonzero $\lambda$.\ \ \


\section{Critical temperature}

Neglecting the background unpaired fermions and modeling our system as a {\it
pure boson gas} of CPs but with temperature-dependent number density
$n_{B}(T)$,
one converts the explicit $T_{c}$-formula (\ref {eq:becmz}) into an {\it
implicit} one by allowing $n_{B}$ to be $T$-dependent.  For $d=2$
(\ref {eq:becmz}) becomes, since $g_{2}(1)\equiv \zeta (2)=\pi ^{2}/6$,
\begin{equation}
T_{c}=\frac{4\sqrt{3}}{\pi ^{3/2}}\frac{\hbar
v_{F}}{k_{B}}\sqrt{n_{B}(T_{c})}.
\label {(11)}
\end{equation}
This requires $n_{B}(T) \equiv N_{B}(T)/L^{2}$ which in turn requires
(\ref {n20t26}), along with $\Delta_{0}(T)$ as determined from (\ref {e13}),
and is given by the expression $2N_{B}(T)/N_{2}=1-N_{20}(T)/N_{2}$.
Solving (\ref {(11)}) self-consistently with $\lambda =1/2$ gives the
remarkably
constant value $T_{c}/T_{F} \simeq 0.004$ over the entire range of
$\nu \equiv \hbar \omega _{D}/E_{F}$ values $0.03 - 0.07$ typical \cite
{Har} of
cuprate superconductors.  On the other hand, the BCS formula
$T_{c}^{BCS}\simeq 1.13\Theta _{D}e^{-1/\lambda }$ with $\lambda =1/2$ gives
$T_{c}/T_{F}$ = 0.005, 0.008 and 0.011 for $\nu$ = 0.03, 0.05 and
0.07, respectively.  Clearly, both sets of predictions are somewhat small
compared with empirical cuprate values of $T_{c}/T_{F}$ that range \cite {Uem}
from $0.01 - 0.1$.

To obtain the exact critical temperature {\it without} neglecting the
background
unpaired fermions, one needs the exact CP excitation energy dispersion
relation
$\varepsilon _{K}(T)\equiv \Delta _{0}(T)-\Delta _{K}(T)$ which is neither
exactly linear in $K$ nor independent of $T$. To determine $\Delta _{K}(T)$ we
need a working equation that generalizes Ref.
\cite{PhysC} for $T>0$ via the new CP eigenvalue equation (\ref{e13}).
Because of symmetry, see Fig. 3, one can restrict the angle $\theta $
to the interval $(0,\pi /2)$ where $k_{1}\geq k_{2}$, i.e., to quadrant I.
Recalling (\ref{bosonenergy}), in $d$-dimensions (\ref{eq:cooper}) becomes
\begin{equation}
1=V\left( \frac{L}{2\pi }\right) ^{d}{\int }^{\prime }d{\bf k}\ \frac{\left[
1-n(\xi _{k_{1}})\right] \left[ 1-n(\xi _{k_{2}})\right] }{\hbar
^{2}(k^{2}-k_{\mu }^{2})/m+\Delta _{K}(T)+\hbar ^{2}K^{2}/4m}.
\label{eq2:cooper}
\end{equation}
Here $k_{\mu }$ is such that $\mu \equiv \hbar ^{2}k_{\mu }^{2}/2m$ and
becomes $k_{F}$ as $T\rightarrow 0$, while $k_{D}$ is such that $\hbar
\omega _{D}\equiv \hbar ^{2}k_{D}^{2}/2m$. The prime on the integral sign
now denotes the restrictions
\begin{eqnarray}
k_{2}^{2}\equiv |{\bf k}-{\textstyle\frac{1}{2}}{\bf K}|^{2} &=&k^{2}-kK\cos
\theta +{\textstyle\frac{1}{4}}K^{2}>k_{\mu }^{2},  \label{2_rest} \\
k_{1}^{2}\equiv |{\bf k}+{\textstyle\frac{1}{2}}{\bf K}|^{2} &=&k^{2}+kK\cos
\theta +{\textstyle\frac{1}{4}}K^{2}<k_{\mu }^{2}+k_{D}^{2}\,.
\label{1_rest}
\end{eqnarray}



\noindent In Fig. 3 the darkest shading corresponds to these (BCS model
interaction) restrictions. The conditions (\ref{2_rest}) and (\ref
{1_rest}) can be studied separately but must be satisfied simultaneously.
If $K<2\sqrt{k_{\mu }^{2}-k_{D}^{2}}$, (\ref{2_rest}) and (\ref{1_rest})
are equivalent to
\begin{equation}
\begin{array}{ccccc}
(k_{\mu }^{2}-{\textstyle\frac{1}{4}}K^{2}\,\sin ^{2}\theta )^{1/2}+{%
\textstyle\frac{1}{2}}K\,\cos \theta & < & k & < & [(k_{\mu }^{2}+k_{D}^{2})-%
{\textstyle\frac{1}{4}}K^{2}\,\sin ^{2}\theta ]^{1/2}-{\textstyle\frac{1}{2}}%
K\,\cos \theta \,.
\end{array}
\label{res_final}
\end{equation}
Note that for $K>\sqrt{k_{\mu }^{2}+k_{D}^{2}}-\sqrt{k_{\mu }^{2}-k_{D}^{2}}$
there exists a minimum value $\theta _{{\rm min}}$ of $\theta $
given by
\begin{equation}
\cos \theta _{{\rm min}}\equiv \frac{k_{D}^{2}}{K\sqrt{2(2k_{\mu
}^{2}+k_{D}^{2})-K^{2}}}\,,  \label{cosmin}
\end{equation}
while $\theta _{{\rm min}}$ = 0 for $K<\sqrt{k_{\mu }^{2}+k_{D}^{2}}-\sqrt{%
k_{\mu }^{2}-k_{D}^{2}}$. We introduce the dimensionless variables

\begin{equation}
\kappa \equiv \frac{K}{2(k_{F}^{2}+k_{D}^{2})^{1/2}}\leq 1,\;\;\;\;\xi
\equiv \frac{k}{k_{F}},\;\;\;\;\tilde{\Delta}_{\kappa }\equiv \frac{\Delta
_{K}}{E_{F}},\;\;\;\;\nu \equiv \frac{\Theta _{D}}{T_{F}}\equiv \frac{%
k_{D}^{2}}{k_{F}^{2}},  \label{eq:4}
\end{equation}
with $k_{B}\Theta _{D}\equiv \hbar \omega _{D}\equiv \hbar ^{2}k_{D}^{2}/2m$
and$\;k_{B}T_{F}\equiv E_{F},$ where $k_{B}$ is Boltzmann's constant. Recall
the $d=2$ constant expression (\ref{gdee}) for $g(\epsilon )$, the
restrictions
(\ref{res_final}), and that for $K\geq 0$ and $T>0$ the step functions in
(\ref{eq:cooper}) $\theta (k_{1,2}-k_{F})\equiv \theta (|{\frac{1}{2}}{\bf K}%
\pm {\bf k}|-k_{F})$ become $[\exp \{-\beta \lbrack \hbar ^{2}({\frac{1}{2}}%
{\bf K}\pm {\bf k})^{2}/2m-\mu (T)]\}+1]^{-1}$---but with $2\epsilon _{k}$
in (\ref{eq:cooper})
replaced by $\epsilon _{k_{1}}+\epsilon _{k_{2}}$, $E_{F}$ by $\mu (T)$
and $\Delta _{K}$ by $\Delta _{K}(T)$.  One finally arrives at a {\it working
equation} for the binding energy $\Delta_{K}(T)$ that generalizes Eq. (18) of
Ref. \cite{PhysC}, namely
\begin{eqnarray}
1 &=&{\frac{4}{\pi }}\lambda \int_{\theta _{min}}^{\pi /2}d\theta \int_{{\xi
}_{min}(\theta )}^{{\xi }_{max}(\theta )}\!\!d{\xi }\,\,{\xi }{\frac{[1+\exp
\{-\tilde{\beta}[{\xi }^{2}+(1+\nu )\kappa ^{2}+2\sqrt{1+\nu }\,\,\kappa {%
\xi }\cos \theta -1]\}]^{-1}}{2\xi ^{2}+2(1+\nu )\kappa ^{2}-2+\tilde{\Delta}%
_{\kappa }(\tilde{T})}}  \nonumber \\
&&\times \lbrack 1+\exp \{-\tilde{\beta}[\xi ^{2}+(1+\nu )\kappa ^{2}-2\sqrt{%
1+\nu }\,\,\kappa \xi \cos \theta -1]\}]^{-1}\,,  \label{weq}
\end{eqnarray}
where $\nu \equiv \hbar \omega _{D}/\mu $,\thinspace\ ${\xi }_{min}(\theta
)\equiv \sqrt{1+\nu }\,\,\kappa \cos \theta +\sqrt{1-(1+\nu )\kappa ^{2}\sin
^{2}\theta }$,\thinspace\ ${\xi }_{max}(\theta )\equiv $\newline
$-\sqrt{1+\nu }\,\,\kappa \cos \theta +\sqrt{(1+\nu )(1-\kappa ^{2}\sin
^{2}\theta )}$ and
\[
\theta _{min}=\left\{
\begin{array}{ll}
& 0\quad \mbox{if}\quad 2\kappa <1-\sqrt{(1-\nu )/(1+\nu )}, \\
& \cos ^{-1}(\nu /\{4\sqrt{1+\nu }\,\,\kappa \sqrt{1+\nu /2-(1+\nu )\kappa
^{2}}\})\ \quad {\rm otherwise}.
\end{array}
\right.
\]
In (\ref{weq}) we have introduced the more general dimensionless quantities $%
{\xi }\equiv k/k_{\mu }$, $\tilde{\Delta}_{\kappa }(\tilde{T})\equiv \Delta
_{K}(T)/\mu $, where $\tilde{T}\equiv k_{B}T/\mu $ or $\tilde{\beta}\equiv
\mu \beta $, and $\kappa \equiv K/2\sqrt{k_{\mu }^{2}+k_{D}^{2}}$.


To obtain the critical temperature from the finite-temperature dispersion
relation, besides solving (\ref{eq2:cooper}) for $\Delta _{K}(T)$, one needs
(\ref{n214}), (\ref{eq:number}), (\ref{N_BK}) and (\ref{muT}). \ At $T=T_{c}$
both $N_{B,0}(T_{c})\simeq 0$ and $\mu _{B}(T_{c})\simeq 0$ so that one gets
the implicit $T_{c}$-equation for the {\it binary mixture gas}

\begin{equation}
1={\frac{\tilde{T}_{c}}{\nu }}\ln \left[ {\frac{1+e^{-\{\tilde{\Delta}_{0}(%
\tilde{T}_{c})/2-\nu \}/\tilde{T}_{c}}}{1+e^{-\{\tilde{\Delta}_{0}(\tilde{T}%
_{c})/2+\nu \}/\tilde{T}_{c}}}}\right] +{\frac{8(1+\nu )}{\nu }}%
\int_{0}^{\kappa _{0}(\tilde{T}_{c})}d\kappa {\frac{\kappa }{e^{[\tilde{\Delta}
_{0}(\tilde{T%
}_{c})-\tilde{\Delta}_{\kappa }(\tilde{T}_{c})]/\tilde{T}_{c}}-1}}.
\label{Tcnum}
\end{equation}
This must be solved numerically for the exact $T_{c}$
for each $\lambda $ and $\nu $ in conjunction with (\ref{e13}) for
$\tilde{\Delta}_{0}(%
\tilde{T})$ and (\ref{weq}) for both $\tilde{\Delta}_{\kappa }(\tilde{T})$ and
$\kappa _{0}(\tilde{T}_{c})$.  Results for $\lambda =1/2$ are shown in Table 1
and Fig. 4 for a range of $\nu $ values typical \cite {Har} of cuprates.

In order to
make this comparison we have taken $T_{\mu }/T_{F}\simeq 1$, a very good
approximation up to the highest temperatures dealt with. \ For example, from
Fig. 4 the highest $T_{c}/T_{F}\simeq 0.14$ already gives $T_{\mu }/T_{F}\simeq
0.9999$ from (\ref{muT}), while for smaller $T_{c}/T_{F}$ the values of $T_{\mu
}/T_{F}$ are even closer to 1. The $T_{c}$ resulting from the exact
dispersion relation for $T=0$ (dot-dashed curve) is somewhat higher than the
exact result (full curve)
but lower than that using the linear approximation for $\Delta_{K}(T)$ (dotted
curve). It is also clear that the
effect of using the exact or linear (in $K$) cases dominates the effect of
the dispersion relation $T$-dependence. For cuprates $d\simeq 2.03$ has been
suggested \cite{wen} to be more realistic as it reflects inter-CuO-layer
couplings but our results in that case would be very similar to those reported
here for $d=2$.

Thus, for $\nu =0.05$ the exact $T_{c}$ is seen to be about 46\% lower
than the heuristic result found in Ref. \cite{cas}, Eqs. (\ref{eq:number}) and
(\ref{eq:chemeqb}). \ It is curious that all results depend very weakly on the
$T$-dependence of the CP binding energy $\Delta _{K}(T)$, in spite of its
being
substantial throughout the temperatures spanned in this paper, as seen in Fig.
2. \ \

We defer study of the condensate fraction $N_{B,0}(T)/N_{B}(T)$ below $T_{c}$
and merely surmise that it may ultimately help
explain the apparent absence \cite{Ruvalds, Martindale} in cuprates of the
Hebel-Slichter peak of nuclear-spin (NMR) relaxation rates {\it vs }%
temperature for $0\leq T\leq T_{c}$. \ Such a peak, originally seen \cite{HS}
in aluminum, is perhaps the most stringent and qualitatively convincing
experimental test of BCS theory (Refs. \cite{Schrieffer}, p. 71 and \cite
{Tinkham}, p. 79 ff). Besides cuprates, it is also absent \cite{Ishiguro} in
several quasi-1D Bechgaard \cite{salts} and in several
quasi-2D (ET) organic salt superconductors. \

\section{Conclusions}


A simple statistical model treating CPs as non-interacting bosons in thermal
and chemical equilibrium with unpaired fermions is proposed. The model gives
rise to a boson number that is strongly coupling- and temperature-dependent.
Since the CP dispersion relation is approximately linear, it exhibits a
Bose-Einstein condensation of zero-CMM pairs at precisely two dimensions.
Exact transition temperatures for the boson-fermion mixture based upon the exact CP dispersion relation are
in reasonable agreement with empirical cuprate data.

Needless to say, further corrections are yet to be included in the present
simple binary mixture boson-fermion model, e.g., i) realistic Fermi surfaces,
ii) Van Hove singularities \cite{Marki} or other means of accounting for
periodic-crystalline effects, as well as iii) the all important $d$-wave
interfermionic interaction, iv) the boson-fermion interaction and v) residual
interbosonic interactions. \ As to the latter, also generally neglected in
BCS theory, if the lowering \cite{London} of $T_{c}$ in liquid $^{4}$He by
about 29\% with respect to the ideal Bose gas BEC $T_{c}$ is any guide,
interbosonic interactions will also lower $T_{c}$\ in a more realistic
picture.
As to the boson-fermion interaction, it is precisely this ingredient that
enabled T.D. Lee and coworkers \cite {TDLee}, and Tolmachev \cite {Tolma} more
generally, to link BCS and BEC through a relation stating that the BE
condensate
fraction is proportional to the (BCS-like) fermionic gap $\Delta(T)$ squared.

\section*{Acknowledgments}

We thank A. Salazar for help with Figure 1.  M.C., A.P. and A.R. are grateful
for partial support from grant
PB98-0124, and M.deLl. from grants PB92-1083 and SAB95-0312, both by DGICYT
(Spain), and PAPIIT (Mexico) IN102198-9 as well as from CONACyT (Mexico)
27828-E. \ R.M.Q. and N.J.D. acknowledge the support of the FRD (South
Africa). \ M.deLl. thanks D.M. Eagles, M. Fortes, O. Rojo, and A.A.
Valladares for numerous discussions; V.V. Tolmachev for extensive
correspondence; A. Polls and R. Escudero for calling his attention to Refs. \cite{Edwards} and \cite{Ruvalds}, respectively; and A.N. Kocharian for reading and commenting the manuscript.

\pagebreak

\pagebreak
\begin{table}
\caption{Critical temperatures $T_{c}/T_{\mu }$ for $\lambda=1/2$ depicted
in Fig. 4 according to (\ref{Tcnum}). The exact result is compared with the
linear-in-$K$ approximation for both $\Delta_{K}(T)$ and $\Delta_{K}(0)$ in
order to test sensitivity of a temperature-dependence of the CP binding energy
for nonzero $K$.}

\begin{center}
\begin{tabular}{|c|c|c|c|}
\hline
$\nu $ & linear approx. with $\Delta _{K}(T)$ & linear approx. with $\Delta
_{K}(0)$ & Exact \\ \hline
0.03 & 0.078 & 0.068 & 0.065 \\ \hline
0.04 & 0.089 & 0.079 & 0.075 \\ \hline
0.05 & 0.100 & 0.088 & 0.084 \\ \hline
0.06 & 0.109 & 0.096 & 0.091 \\ \hline
0.07 & 0.117 & 0.104 & 0.098 \\ \hline
\end{tabular}
\label{Table I}
\end{center}
\end{table}
\pagebreak

{\bf Figure Captions}

{\bf Figure 1.} {Fractional number of pairable fermions that are actually
paired, at three different temperatures, {\it vs}. coupling $\lambda $ for the
present first-principles model (\ref{NB/N2}) (thick curves) and estimated for BCS theory at
$T=0$ as explained below (\ref{NB/N2}) (thin curve). The number of pairable fermions
with the BCS model
interaction used is just (\ref{n214}); {\it all }}of them are actually
paired at $T=0$ in the heuristic BEC model, Ref. \cite{cas} Eq. (23).\ \

{\bf Figure 2.} Temperature dependence of $K=0$ CP binding energy $\Delta
_{0}(T)$ obtained numerically from (\ref{e13}) for $\lambda =1/2$ and $\nu
=0.05$. Note that when $T= \infty $ (\ref{e13}) is analytical for
$\Delta_{0}(\infty)$; the latter then turns out to be about $10^{-8}$, so
that
the curve saturates from above to this value at $T={\infty}$. \ \

{\bf Figure 3.} {Cross-section of overlap \lq \lq volume" in momentum space
(darkest shading) where the tip of the relative wavevector ${\bf k}$ (for two
fermions with wavevectors ${\bf k_1}$ and ${\bf k_2}$) must point for the
attractive BCS model interaction (\ref{inter}) between them to be nonzero and
form a Cooper pair of CMM magnitude $\hbar K$.} \ \

{\bf Figure 4.} {Critical BEC temperature $T_{c}$ in units of $T_{F}$,
resulting for the boson-fermion mixture from (\ref{Tcnum}) for $\lambda =1/2$ for varying $\nu \equiv
\hbar \omega_{D}/\mu \simeq \Theta_{D}/T_{F}$: with no approximations
(full curve); using $\Delta_{K}(T)$ evaluated at $T=0$ (dot-dashed);
using the linear-in-$K$ approximation for $\Delta _{K}(T)$ (dotted). The
dashed
straight line is the BCS formula $T_{c} \simeq
1.13\Theta_{D}e^{1/2\lambda}$ for
$\lambda =1/2$. The very lowest full horizontal line is the solution of the
implicit $T_c$-equation (\ref {(11)}) for the pure unbreakable-boson gas for $\nu =
0.03, 0.05$ and $0.07$. A huge $T_c$ enhancement thus follows {\it from the mere presence} of background unpaired fermions. Cuprate data are taken from Ref. \cite {Uem}.

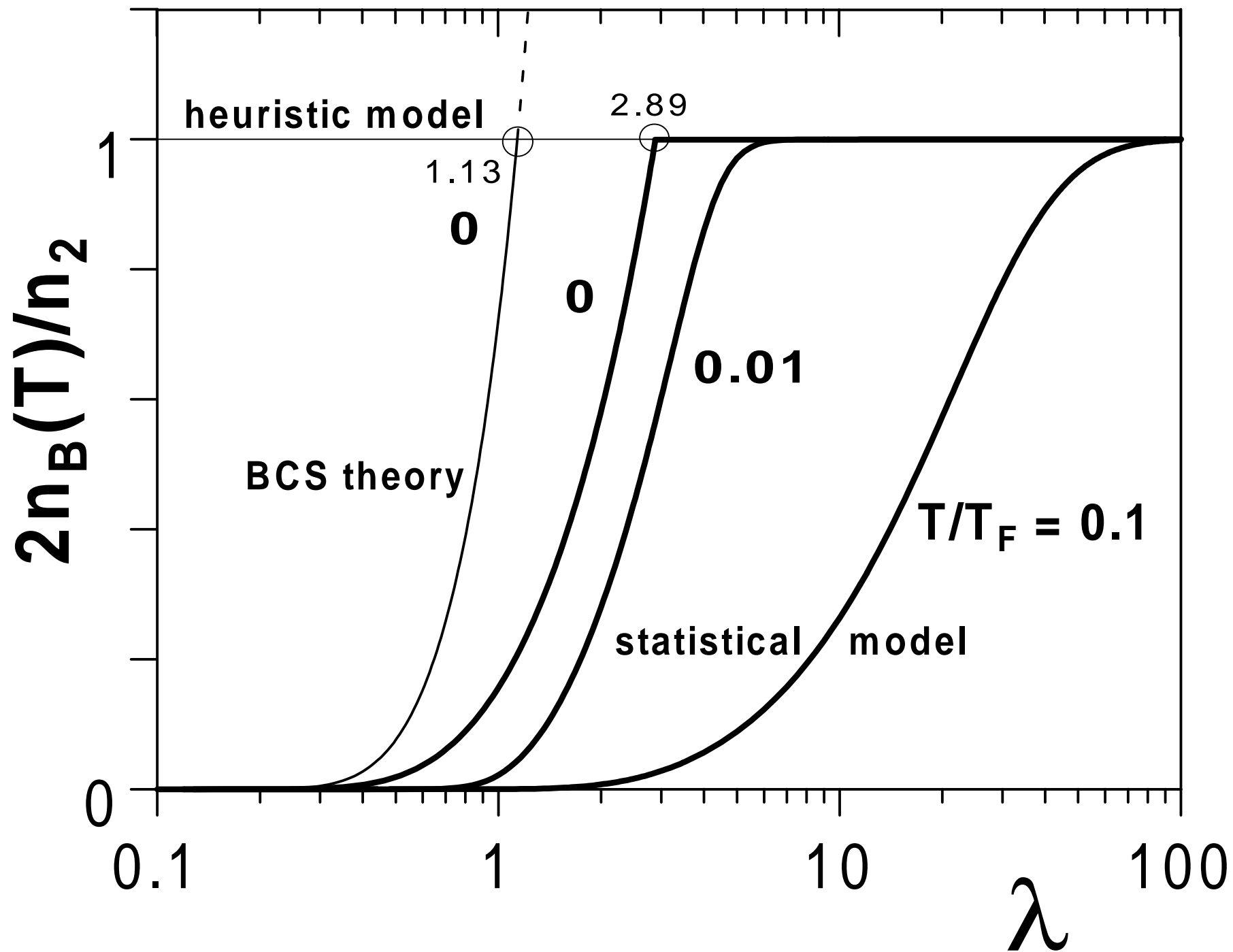
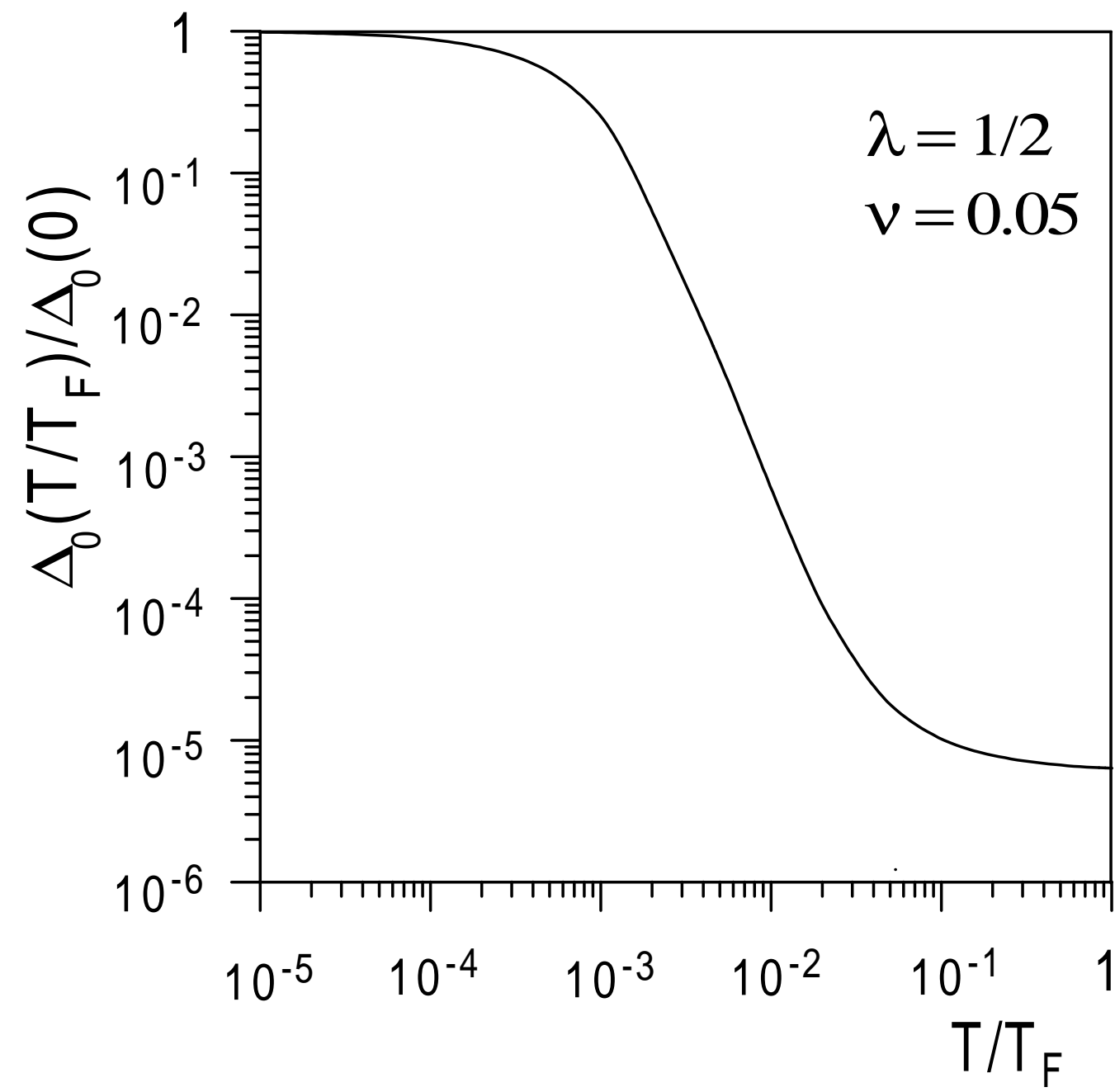
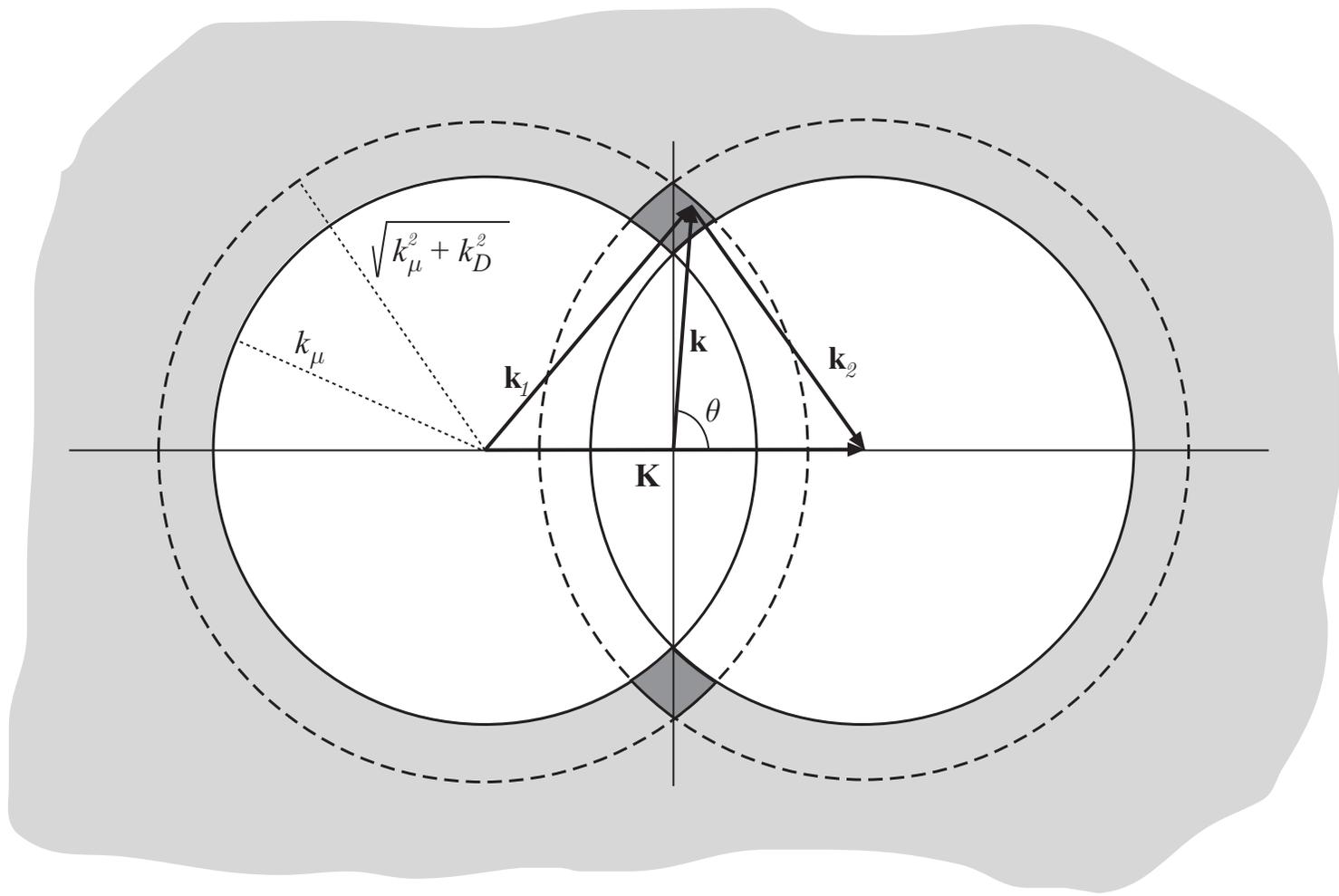
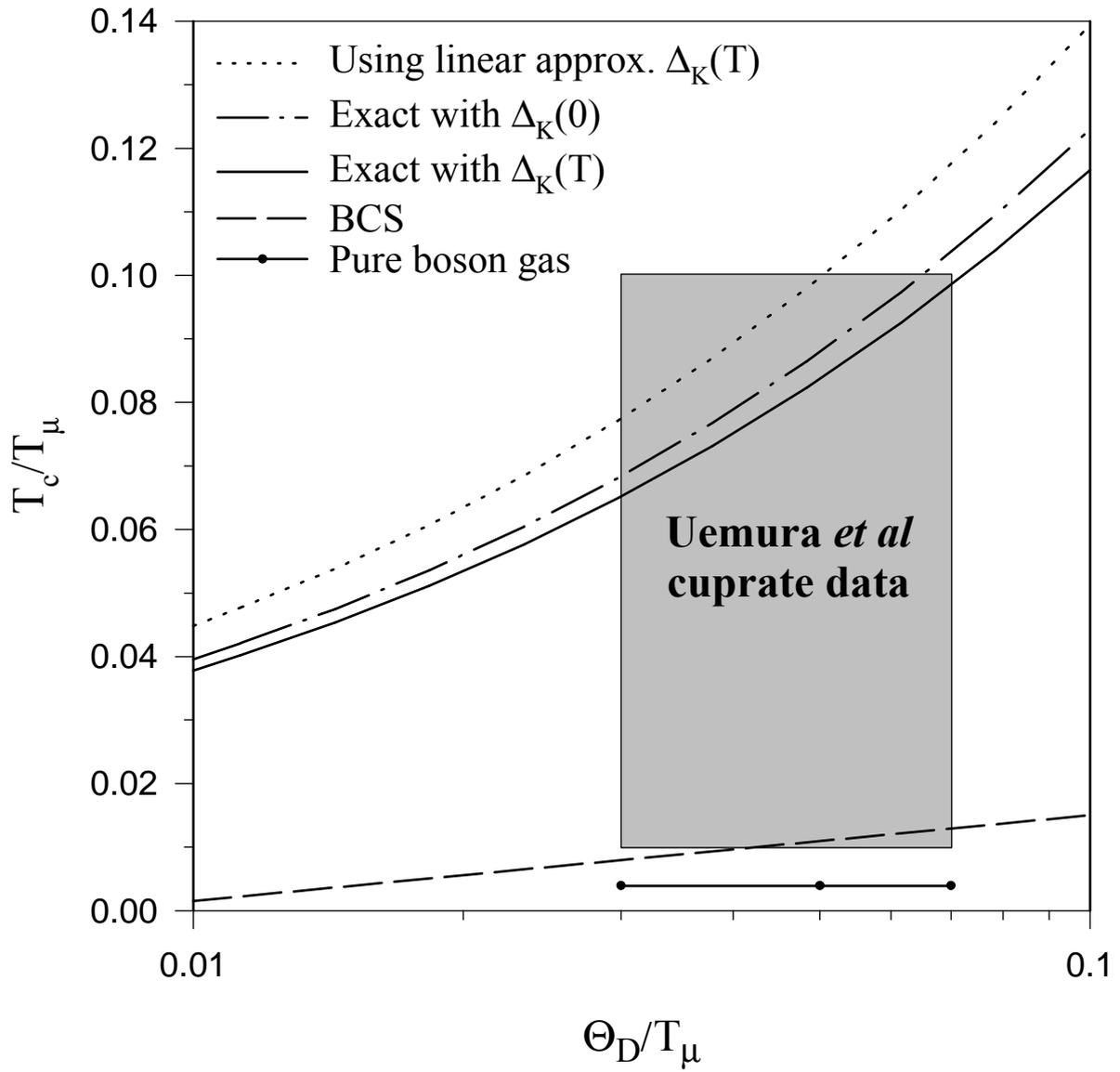
\end{document}